\documentclass[journal=jpccck,manuscript=article]{achemso}

\usepackage[version=3]{mhchem} % Formula subscripts using \ce{}
\usepackage{color}
\usepackage{comment}
\title{Structure and Dynamics of the Quasi-Liquid Layer at the Surface of Ice from Molecular Simulations}

\author{Tanja Kling}
\affiliation{Max-Planck-Institut f\"ur Polymerforschung, Ackermannweg 10, 55128 Mainz, Germany}
\author{Felix Kling}
\affiliation{\textcolor{black}{Institut f\"ur Physikalische Chemie, Duesbergweg 10-14, 55099 Mainz, Germany}} 
\author{Davide Donadio}
\email{ddonadio@ucdavis.edu}
\affiliation[UCD]{Department of Chemistry, University of California Davis, Davis, CA, USA}
 \alsoaffiliation[Ikerbasque]{Ikerbasque, Basque Foundation for Science, Bilbao, Spain}

\abbreviations{IR,NMR,UV}

\begin{document}


\begin{tocentry} 
\includegraphics[width=8.25cm,height=3.60cm]{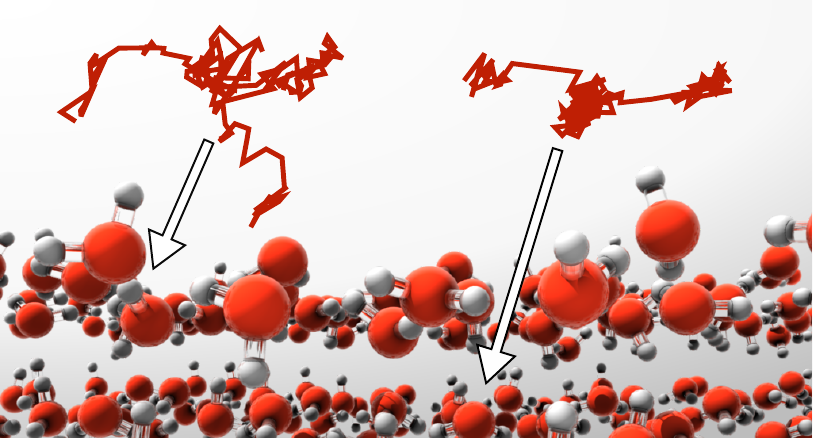}
\end{tocentry}

\begin{abstract}
We characterized the structural and dynamical properties of the quasi-liquid layer (QLL) at the surface of ice by molecular dynamics simulations with a thermodynamically consistent water model. Our simulations show that for three low-index ice surfaces only the outermost molecular layer presents short-range and mid-range disorder and is diffusive. 
The onset temperature for normal diffusion is much higher than the glass temperature of supercooled water, although the diffusivity of the QLL is higher than that of bulk water at the corresponding temperature. 
The underlying subsurface layers impose an ordered template, which produces a regular patterning of the ice/water interface at any temperature, and is responsible for the major differences between QLL and bulk water, especially for what concern the dynamics and the mid-range structure of the hydrogen-bonded network. 
Our work highlights the need of a holistic approach to the characterization of QLL, as a single experimental technique may probe only one specific feature, missing part of the complexity of this fascinating system.
\end{abstract}

\section{Introduction}
Ice surfaces control and catalyze environmentally important chemical reactions \cite{BartelsRausch:2014kj}, such as the formation of ozone depleting species in polar stratospheric clouds \cite{Tolbert:1988fg,Abbatt:1992dg} and adsorption, release and photodissociation of pollutants in snow packs\cite{Beine:2002ft,Chu:2003gb}.
The structure and dynamics of the quasi-liquid layer (QLL) forming at the surface of ice is a key parameter both for the absorption of molecules and for the thermodynamics and kinetics of the reactions that involve adsorbed molecules. Furthermore, QLL dynamics determines the tribological properties of ice, which dictate geological and glaciological processes.

Several experimental probes have been employed to characterize the structure of QLL, including X-ray and electron scattering, ellipsometry and microscopy\cite{Bjorneholm:2016jb,Shultz:2017ee}, often with discrepant results, as for the onset temperature of melting and on the thickness of the QLL \cite{BartelsRausch:2014kj}.
For example, the onset temperature of melting, or formation of the QLL, from photoemission spectroscopy appears to be at 110 K\cite{Nason:1975hp}, from nuclear magnetic resonance at 170 K\cite{Mizuno:1987jp}, from He-atom scattering at 200 K\cite{Suter:2006ca}, from electrical conductivity measurements at 253 K\cite{Bullemer:1966jm} and from ellipsometry at 271 K\cite{Furukawa:1987ct,Nagashima:2016fz}.
QLL thickness and the onset of surface melting are affected by experimental conditions, e.g. vapor pressure\cite{Asakawa:2016cj}, and by the energy exchanged between the sample and more or less invasive probes. A further reason for major discrepancy is that each technique probes different physical quantities, thus assigning the state of the QLL according to different criteria.

Theoretical studies based on molecular dynamics (MD) all agree that up to 1 K below the bulk melting temperature ($T_m$) {the} QLL is less than 1 nm thick, which corresponds to no more than three bilayers for the basal plane of ice I$_h$\cite{Conde:2008bg,Paesani2008:jp,Pfalzgraff:2011de,Benet:2016gu}. Statistical thermodynamic models supported by MD simulations show that, in contrast with first order transitions, surface ice premelting occurs gradually as a function of temperature\cite{Limmer:2014dx,Qiu:2018fe}.
A recent study combining surface sensitive sum frequency generation (SFG) spectroscopy and MD suggested that melting of both the basal and the primary prismatic plane of ice occurs bilayer by bilayer and QLL is intrinsically remarkably thin, i.e. at most two bilayers at 270 K\cite{Sanchez:2016id}. 
However, the nature of the QLL still remains elusive, especially as for the relation between its structure and dynamics, and as for the similarities and differences to bulk supercooled liquid water.  

Here we address the molecular structure and dynamics of the QLL at ice surfaces as a function of temperature by MD simulations. We consider the two bilayer low-index basal (0001) and first prismatic (10$\bar{1}$0) surfaces, as well as the single-layer secondary prismatic surface ($\bar{1}$2$\bar{1}$0). 
MD simulations highlight substantial differences between two-dimensional QLL and bulk water, stemming from reduced dimensionality and the template effect of the underlying bulk ice layers. Furthermore we show that a meaningful characterization of the ice surfaces originates from concurrently taking into account both local and mid-range structural properties, as well as dynamical properties.

\section{Methods and models}

Models for molecular simulations of water have been developed in very large number since the early 1980s with varying complexity, from rigid-molecule point-charge models\cite{Jorgensen:1983gg,Berendsen:1987gc} which allow for inexpensive large-scale simulations, to highly transferable polarizable models fitted on high-level {\sl ab initio} data\cite{Babin:2012kv}, which enable accurate calculations of infrared and SFG spectra\cite{Medders:2015dn,Medders:2016gz}.
In this study we address a number of simulations of medium-size systems for relatively long timescales: the total simulation time reaches 1 $\mu$s, which is hardly affordable to {\sl ab initio} MD. Hence we decided to opt for a rigid-molecule point-charge model. 

Among the rigid point-charge models, the class of four-sites ``TIP4P" models\cite{doi:10.1063/1.445869} was found to reproduce qualitatively the phase diagram of water and of the molecular phases of ice\cite{Sanz:2004gh}. This feature is related to the ratio between the dipole and quadruple in the model, and it is retained upon modifying the specific parameters of the model, i.e. charges and van der Waals interactions\cite{Abascal:2007hh}. Among the several parameterizations of the original TIP4P, the one denominated TIP4P/ice reproduces the correct melting temperature and the $dP/dT$ negative slope of the coexistence curve\cite{Abascal:2005jl}. While the thermodynamics of the solid/liquid phase transition is well reproduced by TIP4P/ice, we need to point out that this model significantly underestimate the diffusion coefficient of water\cite{Vega:2009fr}.

Simulations were carried out for orthorhombic ice I$_h$ slab models made of 1536 H$_2$O molecules in a \textcolor{black}{supercell with three-dimensional periodic boundary conditions. Long range electrostatics is computed using the smooth particle-mesh Ewald method}\cite{Essmann:1998gi}. For all the temperatures considered we ran simulations at least 50 ns long, while closer to the melting point our simulation time extends to 200 ns. 
To construct the slab models we prepared bulk models of hexagonal ice, in which proton disorder is generated by a Monte Carlo algorithm that minimizes the total dipole of the simulation cell\cite{Grishina:2004jf}.
To take into account thermal expansion, we ran simulations of these bulk ice models in the constant pressure canonical ensemble\cite{Parrinello:1981vy,Nose:1984wa} for each temperature in the range between 200 K and 270 K, at a pressure of 0 bar. These well-equilibrated models are then cleaved by introducing a vacuum layer of about 20 \AA . Such thickness of the vacuum layer is sufficient to prevent interactions between the two juxtaposed surfaces through periodic boundary conditions. In fact a buffer of about ten molecular diameters already provides reliable estimates for the surface tension of the water vapor/liquid interface\cite{Conde:2008bg,doi:10.1063/1.2715577}
The slabs are equilibrated for 10 ns in the canonical ensemble, enforced by stochastic velocity rescaling\cite{Bussi:2007cs} with a coupling constant of 2 ps. The same thermostat is used to perform production runs in the canonical ensemble. In all MD simulations the equations of motion are integrated with a time step of 1 fs. 

\section{Results and discussion}

\subsection{Local and mid-range structure}

\begin{figure}
\includegraphics[width=8.cm]{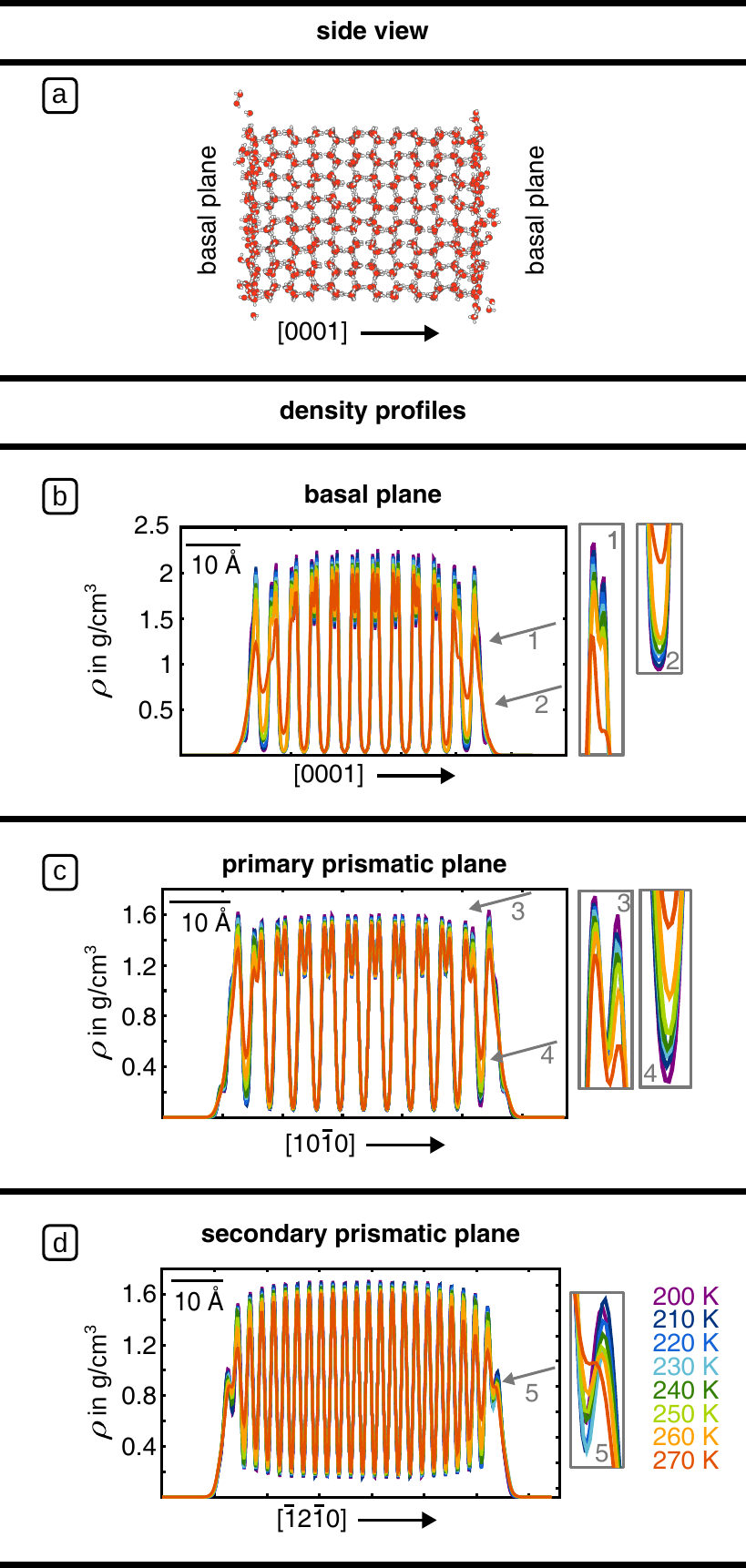}
  \caption{Side view of a basal slab at 200 K after 10 ns equilibration (a).
Density profiles as a function of temperature for the basal (b), primary prismatic (c), and secondary prismatic plane (d) of hexagonal ice, illustrating the bilayer and monolayer structures, respectively. Side panels show close-up of regions in which prominent changes happen as the temperature is raised from 200 to 270 K.} 
  \label{fig:densprofiles}
\end{figure}

% Density and g(r): refer to PNAS. 
As shown in previous works\cite{Paesani2008:jp,Conde:2008bg,Pfalzgraff:2011de,Sanchez:2016id}, computing the mass density profile in the direction perpendicular to the surfaces provides a first-glance indication of the surface melting behavior (Figure~\ref{fig:densprofiles}). In the bulk region density profiles exhibit the fingerprints of bilayers for basal and first prism plane and equally spaced monolayers for the secondary prismatic plane. The most external layers at the ice/vacuum interface are smeared. For (0001) and (10$\bar{1}$0) surfaces the double peak typical of bilayers is smeared even at the lowest temperature considered (Figure~\ref{fig:densprofiles}, insets 1-4). This finding suggests that as--cleaved bilayers rapidly rearrange into different structures in order to minimize the number of broken hydrogen bonds. Similar behavior was observed for ice at weakly hydrophilic interfaces\cite{Andreussi:2006er}. Such low temperature structures have been recently characterized by vibrational spectroscopy, also supporting the occurrence of a disordered layer at $\sim$200 K\cite{Bisson:2013fh,Smit:2017kl}. At lower temperatures the crystalline structure may be retained, and surprisingly proton-ordered surface turn out energetically favored\cite{Buch:2008ij}, \textcolor{black}{as observed by heterodyne-detected SFG\cite{Nojima:2017ei}.} 

As the temperature increases the maximum of the outermost density peaks decreases and the valley separating them from the second layer increases. The close-up panels in Figure~\ref{fig:densprofiles} show that for the single-layer surface this change occurs gradually, while bilayer surfaces exhibit a faster transition from 260 to 270 K. In the case of the basal plane this change is accompanied by a smearing of the bilayer fingerprint in the subsurface layer, which marks the onset of disordering of the second bilayer. 
These results, substantiated by vibrational spectroscopy measurements, were interpreted as discrete bilayer-by-bilayer melting as a function of temperature\cite{Sanchez:2016id}. Yet, the ``molten" state of the second bilayer needs further analysis, as we will see in the following. 

In general, the layering is preserved throughout the whole range of temperatures considered, which makes it possible to perform layer-by-layer structural analysis. In addition, due to the symmetry of the slabs, we can average over the two sides of the slabs for better statistics. 
Calculations of the layer-resolved radial distribution functions as a function of temperature, compared to those of reference models of ice and supercooled water at the corresponding temperatures, suggest that the structure of the outermost layers, which appear to be molten from the density profiles, is akin to that of water, while the inner layers remain ice--like (Figure S1 in Supporting Information -- SI). 

\begin{figure}
  \includegraphics[width=8truecm]{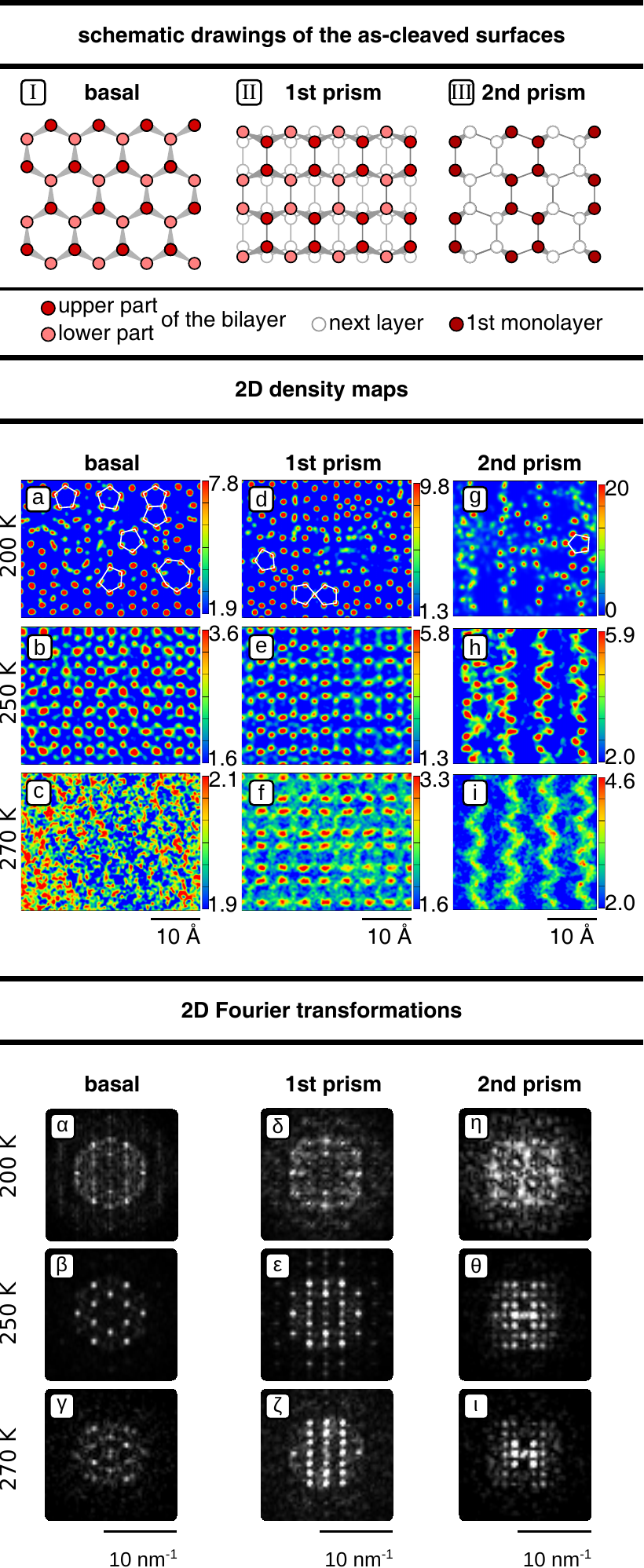}
  \caption{Schematic drawings of the as--cleaved three low-index surfaces ({I}--{III}). Two-dimensional density maps of the oxygen atoms belonging to the first layer of the basal ({a}--{c}), primary prismatic ({d}--{f}), and secondary prismatic plane ({g}--{i}) at three selected temperatures. The color scale represents the probability (in $10^{-5}$ per pixel) to find a single atom at a pixel. In addition, the respective two-dimensional Fourier transforms of the density maps illustrate the periodicities ($\alpha$--$\iota$).}
  \label{fig:2Ddensitymapspart}
\end{figure}
Electron diffraction, He-atom and grazing X-ray scattering are powerful techniques to probe surface crystal structures and have been employed to characterize ice surfaces\cite{Suter:2006ca,Dosch:1995kr,Materer:1997co}. Since these experiments probe surface topology, 
we computed the time-averaged two-dimensional oxygen density maps of the layers undergoing melting or disordering. 
To connect to scattering experiments we also computed the two-dimensional Fourier transform of the density maps, which would resemble diffraction patterns. Examples at low (200 K), intermediate (250 K) and high temperature (270 K) are shown in Figure~\ref{fig:2Ddensitymapspart} (Density maps at all temperatures are shown in Figure S2, and density maps for the subsurface layer are shown in Figure S3). 
Density maps at low temperature show that the hexagonal pattern of the as-cleaved surfaces is broken, and the molecules relax into different structures forming 5-fold rings as well as larger rings, as highlighted in Figure~\ref{fig:2Ddensitymapspart}a,d and g. Ring size distributions will be discussed below. 
The sharp features in the map indicate that after a first rapid rearrangement, which occurs in the 10 ns equilibration, surface molecules hardly diffuse, leaving the disordered structure unchanged over the simulations timescale (40 ns). 
In spite of the presence of topological defects, the underlying symmetry of all three surfaces is still visible in the 2D-FT, although the patterns are blurred due the reduced symmetry (Figure~\ref{fig:2Ddensitymapspart}$\alpha - \iota$). 

Surprisingly, increasing the temperature to 250 K disrupts the topological defects and produces more ordered surface patterns for all the surfaces, accompanied by sharper diffraction patterns. Observing single surface snapshots it becomes evident that the nature of the restored surface order is dynamic (Figure S4). In fact single configurations appear disordered, but molecules diffuse on an underlying crystalline layer, which provides a regular template corresponding to the as-cleaved surface structure for all three surfaces (Figure~\ref{fig:2Ddensitymapspart} {I}--{III}). 
The occurrence of surface patterning provides an indirect estimate of the onset of surface mobility at a temperature between 240 and 250 K, which is approximately the same for all the surfaces (Figure~\ref{fig:2Ddensitymapspart}b,e,h).

The ordered patterns on the primary and secondary prism planes are retained up to 270 K.
The basal plane, instead, does not exhibit obvious patterning at 270 K, but the 2D-FT of the image still yields the pattern corresponding to a hexagonal layer, indicating that the template effect is still present even when the second bilayer below the surface is disordered. 
These results indicate that the structure of the QLL differs significantly from that of bulk water, even when the local ordering of the molten layers, resembles that of water. Such differences stem from the patterning effect of the underlying bulk crystalline layers, which affects the planar structure of the liquid. 

% FIGURE: Ring statistics
\begin{figure}
  \includegraphics[width=8.250cm]{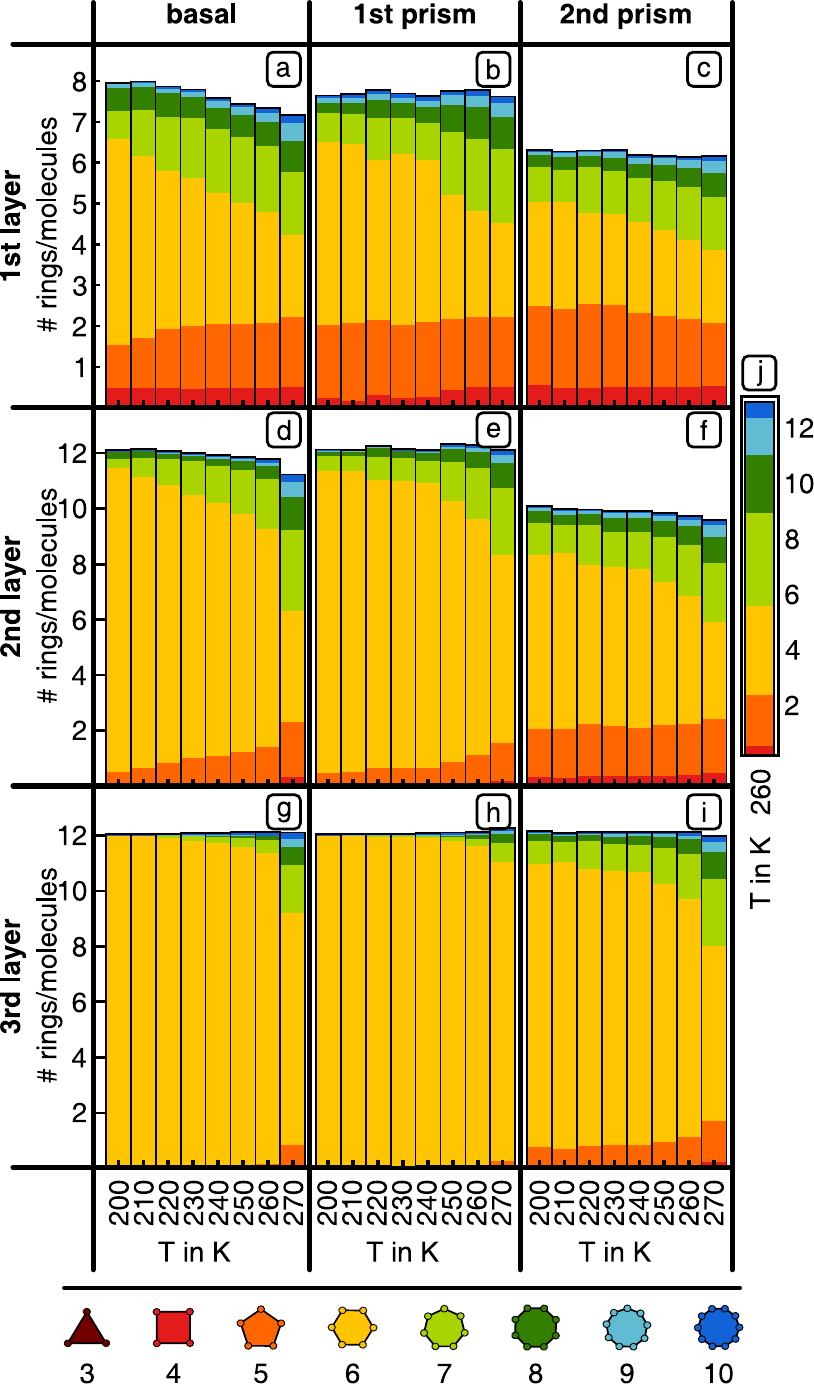}
  \caption{Layer-resolved rings distributions normalized over the number of water molecules for the surface layer and for the two layers beneath. The rings distribution of supercooled water at \textcolor{black}{260 K} is provided for reference. The rings statistics of supercooled water does not change significantly as a function temperature\cite{Belch:1987fi}.}
  \label{fig:3rings}
\end{figure}
Both in bulk water and pre-melted ice surfaces, water molecules arrange in a hydrogen-bonded random tetrahedral network (RTN)\cite{Sciortino:1991hh}. The distribution of rings is an effective tool to characterize the mid-range order of RTNs, such as liquid water and amorphous ices\cite{Belch:1987fi,Martonak:2005dw}. Here we calculated layer-resolved primitive rings distributions\cite{Franzblau:1991wy} of ice slabs as a function of temperature (Figure~\ref{fig:3rings}).
To define hydrogen bonding we used a geometric criterion, according to which two water molecules are considered hydrogen-bonded when the O--H distance is smaller than 2.5 \AA\ and the OHO angle is smaller than 135$^o$.
According to the definition of primitive ring, bulk crystalline ice I$_h$ consists exclusively of 6-membered rings.
Hence, the presence of rings of different size provides an immediate measure of the disordering of the surface layers.
The substantial number of 4, 5, 7 and 8-membered rings in the outermost ice layer at 200 K is a measure of the spontaneous disordering that is present at the surface of ice at low temperature. 
As temperature increases also disorder in the first and, eventually second layer, increases, with progressive reduction of the number of 6-membered rings. The final distributions for the first layers are similar for all the surfaces and feature an excess of 5 and 7-membered rings. It is worth noting that disordering of the second layer for the basal plane occurs at first gradually, but the number of 6-membered rings shows a sudden jump from 260 K to 270 K {(Figure~\ref{fig:3rings}d)} indicating a sharp disordering transition at this temperature. {The} transition is much smoother for the other surfaces.

Although a direct comparison between two-dimensional confined systems, such as QLL, and a three-dimensional bulk system may be impervious{,} we still provide the rings distribution of supercooled water at 300 K for reference {(Figure~\ref{fig:3rings}j)}. The main difference between QLL and bulk water lies in the excess of 5-membered rings and the much lower number of larger 8, 9 and 10-membered rings. 
The hexagonal patterning discussed above and the differences in the rings distribution suggest that the mid-range structure of QLL is substantially different from that of bulk water {at the same temperature}. 
%I cite a paper that shows that the rings distribution of water does not change much with temperature: see caption.
% Patterning may be seen by electron/atom scattering
% Disordering may be probed by NMR and photoelectron spectroscopy

\subsection{Dynamics}

To characterize surface melting, the structural analysis discussed in the previous section needs to be complemented by the study of dynamics. 
Two types of mechanisms take place in the QLL: in-plane diffusion and hopping among different layers\cite{Gladich:2011jn,Pfalzgraff:2011de}.
A detailed characterization of the interlayer molecule exchange mechanism can be found in \citealt{Pfalzgraff:2011de}: our simulations confirm previous results and will not be discussed in this paper.
Here we focus on a layer-resolved analysis of the in-plane diffusion in the QLL. 

To this aim, we calculate layer-resolved mean square displacement (MSD), which is defined as $MSD(t)=\left\langle (x(t)-x(t_0))^2\right\rangle$ averaged over the molecules that belong to a specific layer. Because of the rapid molecule exchange between the layers, while computing $MSD(t)$ we made sure to trace whether molecules stay in the the same layer for the whole time $t$. 
In general $MSD(t)\propto t^\eta$. For normal (Fickian) diffusion, for an $n-$dimensional system, the MSD is linear with time ($\eta=1$) and the slope is given by the diffusion coefficient $D$ as $MSD(t)=2nDt$\cite{Einstein1905,Smoluchowski}. 
Over time scales shorter than molecular collisions ballistic motion occurs with $\eta=2$. 
When the exponent $\eta$ is smaller than 1 then dynamics is defined {\it subdiffusive} or {\it glassy}\cite{glassydiff}. 

The patterning observed in the surface density maps (Figure~\ref{fig:2Ddensitymapspart}) could imply anisotropic diffusion, especially for the secondary prismatic plane, where dimer rows remain clearly visible up to 270 K.
In fact, former simulations suggested anisotropic in-plane diffusion for the primary prismatic plane, but not for the basal plane\cite{Gladich:2011jn}. 
\textcolor{black}{To test the possible anisotropy of in plane diffusion we have computed a histogram of the \textcolor{black}{displacement vectors of} the water molecules in the outermost layers in 1 ns, averaged over the whole trajectories at low (200 K) intermediate (250 K) and high temperature (270 K). These plots do not show any preferential diffusion direction (see Figure S5). Furthermore,  the statistical differences in the MSD between the two symmetric surfaces of the same slab are larger than the differences of the MSD computed along the in plane low-index crystallographic directions for all the surfaces (see Figure S6 and S7).
We can then conclude that in our simulations diffusion does not display any significant in-plane anisotropy for any surface, in agreement with \citealt{Bolton:2000cl}.}
% FIGURE: MSD and exponents 
\begin{figure}
  \includegraphics[width=17.099cm]{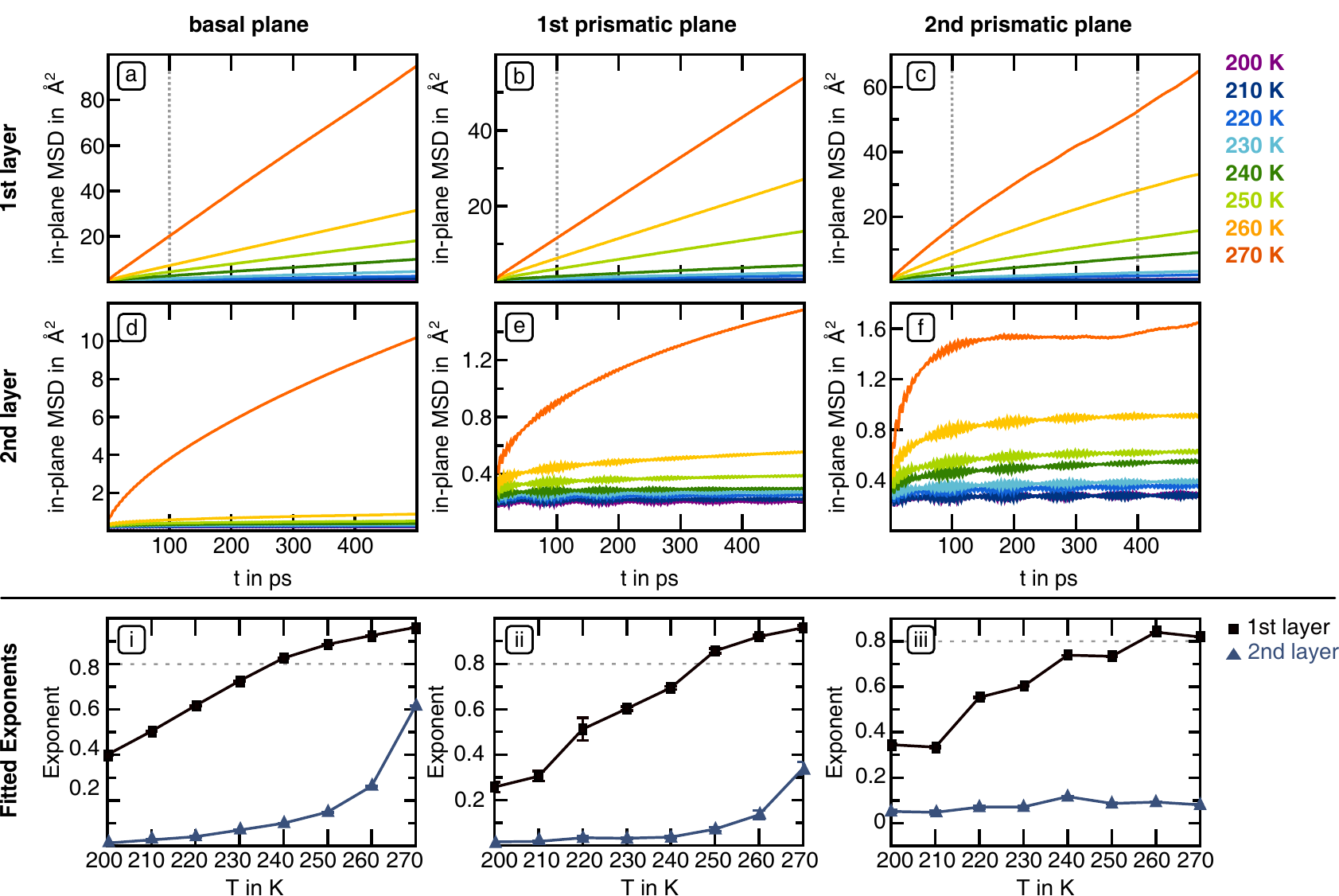}
  \caption{Layer-resolved in-plane mean square displacement for the outermost surface layers (a--c) and the subsurface layers (d--f) of the three ice slabs considered: basal plane (a,d), primary prismatic (b,e) and secondary prismatic (c,f). Panels (i--iii) report the exponents $\eta$ fitted from $MSD(t)\propto t^\eta$ as a function of temperature.}
  \label{fig:4inplaneMSD}
\end{figure}

Figure~\ref{fig:4inplaneMSD} shows the MSD of the surface and the subsurface layer for each ice slab considered. The first row of panels (Figure~\ref{fig:4inplaneMSD}a-c) suggests that the outermost layer for each surface is diffusive at high enough temperatures. 
In contrast, we observe negligible MSD, far below a square molecular diameter over $t=0.5$ ns, for all the subsurface layers at any temperature considered. This implies that even if the second layers appear disordered from the structural analysis, the mobility of the molecules is too low to consider them liquid.

\begin{figure}
  \includegraphics[width=7.877cm]{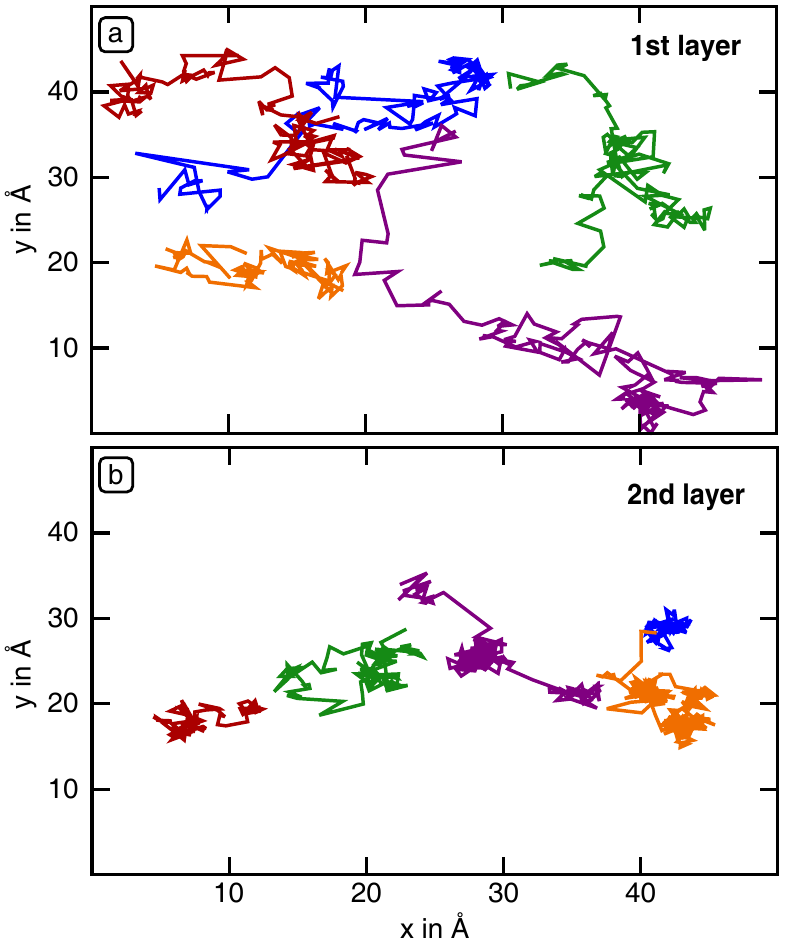}
  \caption{Trajectories of five molecules in the outermost (a) and in the subsurface layer (b) of the basal plane at 270 K.}
  \label{fig:6hopping}
\end{figure}
Figure~\ref{fig:6hopping} provides a pictorial representation of the differences between the diffusive dynamics of the outermost layer and the subdiffusive dynamics of the subsurface layer of the basal plane at 270 K. The molecules in the outermost layer display limited tendency to be trapped in specific sites and undergo normal diffusion, while those in the subsurface layer diffuse by hopping from one cage site to the next. Combining this information with the 2D density maps discussed in the previous section {(Figure S3)} suggests that these cage sites correspond to the crystallographic sites of ice I$_h$. 

The onset temperature of diffusivity for the top layer is defined quantitatively by fitting the exponent $\eta$ on the $MSD(t)$ curves. 
Considering also the structural features of the surface layer, we chose as an arbitrary cutoff for normal diffusion $\eta>0.8$. According to this definition, the onset temperature of diffusivity is between 230 and 240 K for the basal plane, between 240 and 250 K for the first prism plane and between 250 and 260 K for the secondary prism plane. 
Below these temperatures surface molecules display limited mobility and $\eta$ decreases for decreasing temperatures. 
Remarkably, we do not find a sharp drop of $\eta$ as a function of temperature, as it would happen in the case of a first order transition, e.g. crystallization, but rather $\eta$ decreases gradually as for the case of a glass transitions\cite{Charbonneau:2017hv}. However, for supercooled bulk water the glass transition is predicted to occur at much lower temperature (130--150 K)\cite{Mishima:1998hf, Debenedetti:2003gn}, thus suggesting that the mechanisms that lead to dynamical arrest are substantially different. While surface molecules have the freedom to move vertically at the liquid/vacuum interface of the QLL, they also undergo a pinning effect from the underlying crystalline subsurface layer, which turns out the decisive factor to limit molecular mobility and to cause glassy dynamics. 
It is necessary to point out that in the subdiffusive transport regime a linear fit of MSD(t) over timescales of few tens of picoseconds is unjustified to characterize the dynamics of the ice-air interface\cite{Weber:2018ba}. 
A more detailed characterization of the glassy dynamics of the QLL would require the analysis of structural and dynamical correlations over longer simulation times\cite{Appignanesi:2006hs,Appignanesi:2009im}, also considering that normal diffusivity may set in at time scales not accessible to our MD simulations\cite{Charbonneau:2017hv}. However, an important issue lies in the limited latency time of molecules within the surface layer, which may make normal diffusion not observable even in longer simulations.

\begin{figure}
  \includegraphics[width=8.059cm]{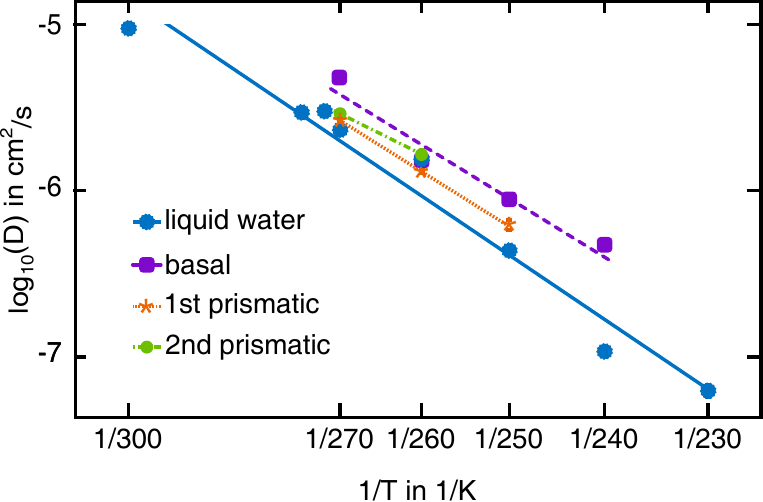}
  \caption{\textcolor{black}{Arrhenius plot of the in-plane diffusion coefficients for the top layers of the three simulated ice surfaces and  of the diffusion coefficients of bulk supercooled water modeled with the same TIP4P/ice forcefield.}
 % \textbf{\textcolor{black}{\textcolor{blue}{Remove this:}In plane two-dimensional diffusivity of the outermost surface layers of ice as a function of temperature, compared to the three-dimensional diffusion coefficient of supercooled water modeled with the same TIP4P/ice forcefield.}} 
  Error bars are smaller than the size of the symbols.}
  \label{fig:5diffusivity}
\end{figure}

For the range of temperatures, at which the surface layers are diffusive, \textcolor{black}{we computed the in-plane diffusion coefficients and compared them to the diffusion coefficients of bulk supercooled water computed with the same TIP4P/ice force field (Figure ~\ref{fig:5diffusivity}). QLL diffusion coefficients turn out higher than that of bulk water at the corresponding temperature. This is readily explained by the lower connectivity of the hydrogen-bonded network due to the interface with air.} 
%\textbf{\textcolor{blue}{Remove this:}\textcolor{black}{we computed their two-dimensional self-diffusion coefficients  and we compare them to the three-dimensional diffusion coefficients of supercooled water, computed with the TIP4P/ice force field (Figure ~\ref{fig:5diffusivity}). QLL diffusion coefficients turn out higher than that of bulk water at the corresponding temperature.With the caveat that we are comparing different quantities, i.e. 2D and 3D diffusion coefficients, we argue that the diffusivity of the surface molecules is enhanced by the lower connectivity of the hydrogen-bonded network due to the interface with air.}} 
Surprisingly, in the diffusive regime, the interaction with the underlying subsurface and bulk layers is not sufficient to limit the mobility of the surface molecules.  

The kinetics of normal diffusion can be interpreted as an Arrhenius process at the molecular level, from which the temperature dependence of the self-diffusion coefficient $D(T)$ is expressed as: 
\begin{equation}
\label{eq:Arrhenius}
\mathrm{ln}(D(T)) =  \mathrm{ln}(D_0) - \frac{E_{\text{A}}}{\mathrm{k_B}\cdot T}. 
\end{equation}
$E_{\text{A}}$ is the activation energy of the diffusion process, $D_0$ is a temperature-independent constant, often referred to as the pre-exponential factor and $k_B$ is Boltzmann's constant. From the self-diffusion coefficients in Figure~\ref{fig:5diffusivity} we obtain activation energies $E_{\text{A}}=40.5\pm0.7$ kJ/mol for the basal plane and $E_{\text{A}}=40.8\pm 2.4$ kJ/mol for the primary prismatic plane. $E_A$ is lower for the secondary prismatic plane ($E_{\text{A}}=33.4\pm 4$ kJ/mol), but it is necessary to point out that this estimate is obtained with a statistically limited set of data, as the surface is diffusive only at 260 and 270 K. 
In general the activation energies for QLL are slightly smaller than those obtained for bulk water using the same force field (44.4 kJ/mol)\cite{Espinosa:2014di}.
These estimates for QLL disagree with recent tribological measurements that give $E_{\text{A}}\sim 11.5$ kJ/mol\cite{Weber:2018ba}. While discrepancies may arise from the choice of the empirical model, it must be considered that the Arrhenius activation energy of TIP4P/ice bulk water is in good agreement with experimental estimates\cite{Price:1999ie,Espinosa:2014di}. Given the structural and dynamical similarities between bulk water and QLL pointed out so far, it is very unlikely that the diffusion barrier would change by so much as a factor four. 

\section{Conclusions}

In summary, we have characterized the QLL at three different ice surfaces computing several structural and dynamical properties at the molecular level. 
Short-range and medium range structural features are intertwined with the molecular dynamics in a complex way, which would mystify the interpretation of experiments that probe only a subset of such properties. 
\textcolor{black}{The vibrational properties of the QLL probed by SFG resemble those of supercooled liquid water~\cite{Smit:2017jq}, but the medium-range structure and the dynamics of ice surfaces turn out very different from those of bulk water.} 
At low temperature ($\sim$200 K) the surface of ice is disordered as in a liquid, but the molecular mobility is extremely limited, as in an amorphous system, well below the glass transition. Close to the melting point surfaces are slightly more diffusive than bulk water, but they still retain a dynamic ordering, which originates from the underlying bulk ice template: diffusivity and ordering may be probed by different experiments, such as scanning probes for the former and electron or atom scattering for the latter. 
In any case, at 270 K and below only the outermost surface layer is diffusive and can be considered liquid. 
Our simulations reconcile different views of the surface of ice under the umbrella of a molecular picture\cite{Michaelides:2017ba}, however to properly address the interpretation of specific experiments it turns out necessary to calculate the corresponding experimental observables by means of transferable approaches, such as quantum chemistry or quantum-chemically accurate empirical potentials\cite{Medders:2016gz}.

%% classes. This should be given within the "acknowledgement"
\begin{acknowledgement}
This work is based on material from TK's doctoral dissertation\cite{Kling.100001923}.
The authors thank Burkhard D\"unweg for helpful comments. 
TK acknowledges financial support from the Max Planck Graduate Center. 
This material is partly based upon work supported by the National Science Foundation under Grant No. 1806210 and by the Hellman Fellows Fund.

\end{acknowledgement}

\begin{suppinfo}

Layer-resolved radial distribution functions, surface snapshots, two-dimensional density maps at any temperature {considered}, two-dimensional density maps for the subsurface layers, two-dimensional diffusion plots, {in-plane direction}-resolved mean square displacements, and schematic representation of the surface planes are provided as supporting information. 

\end{suppinfo}

%% Notice that the class file automatically \bibliographystyle
\bibliography{theBiggestBlackestLibrary}

\end{document}